\documentstyle[12pt]{article}

\catcode`\@=11 
 
\@addtoreset{equation}{section} 
\@addtoreset{footnote}{section} 
\catcode`\@=12 

\newtheorem{proposition}{Proposition}
\newtheorem{corollary}{Corollary}
\newcommand{\rd}{\partial}
\newcommand{\dfrac}[2]{ \frac{\displaystyle #1}{\displaystyle #2} }

\newcommand{\Res}{\mathop{\mbox{\rm Res}}}
\newcommand{\diag}{\mathop{\mbox{\rm diag}}}
\newcommand{\Tr}{\mathop{\mbox{\rm Tr}}}

\newcommand{\gl}{\mbox{gl}}
\newcommand{\GL}{\mbox{GL}}
\newcommand{\bfC}{{\bf C}}
\newcommand{\bfZ}{{\bf Z}}
\newcommand{\calA}{{\cal A}}
\newcommand{\calB}{{\cal B}}
\newcommand{\calF}{{\cal F}}
\newcommand{\calO}{{\cal O}}
\newcommand{\calT}{{\cal T}}
\newcommand{\Yhat}{\hat{Y}}
\newcommand{\mutilde}{\tilde{\mu}}
\newcommand{\omegatilde}{\tilde{\omega}}
\newcommand{\Omegatilde}{\tilde{\Omega}}
\newcommand{\Ctilde}{\tilde{C}}
\newcommand{\Ftilde}{\tilde{F}}
\begin{document}

\title{Spectral Curves and Whitham Equations in 
Isomonodromic Problems of Schlesinger Type}
\author{Kanehisa Takasaki 
\thanks{Partly supported by the Grant-in-Aid for Scientific 
Researches, Priority Area 231 ``Infinite Analysis'', 
the Ministry of Education, Science and Culture, Japan}
\\
{\normalsize Department of Fundamental Sciences, Kyoto University}\\
{\normalsize Yoshida, Sakyo-ku, Kyoto 606, Japan}\\
{\normalsize E-mail: \tt takasaki@yukawa.kyoto-u.ac.jp}
}
\date{}
\maketitle

\begin{abstract}
It has been known since the beginning of this century that 
isomonodromic problems --- typically the Painlev\'e 
transcendents --- in a suitable asymptotic region look like 
a kind of ``modulation'' of isospectral problem. This 
connection between isomonodromic and isospectral problems 
is reconsidered here in the light of recent studies related 
to the Seiberg-Witten solutions of $N = 2$ supersymmetric 
gauge theories.  A general machinary is illustrated in a 
typical isomonodromic problem, namely the Schlesinger equation, 
which is reformulated to include a small parameter $\epsilon$. 
In the small-$\epsilon$ limit, solutions of this isomonodromic 
problem are expected to behave as a slowly modulated 
finite-gap solution of an isospectral problem. The modulation 
is caused by slow deformations of the spectral curve of the 
finite-gap solution. A modulation equation of this slow 
dynamics is derived by a heuristic method.  An inverse 
period map of Seiberg-Witten type turns out to give general 
solutions of this modulation equation. This construction of 
general solution also reveals the existence of deformations 
of Seiberg-Witten type on the same moduli space of spectral 
curves. A prepotential is also constructed in the same way as
the prepotential of the Seiberg-Witten theory. 
\end{abstract}
\bigskip

\begin{flushleft}
KUCP-0105\\
solv-int/9704004
\end{flushleft}
\newpage

\section{Introduction}

The notion of ``isomonodromic deformations'' was first 
discovered by R. Fuchs \cite{bib:RFuchs} in 1907 as 
a new interpretation of the  6th Painlev\'e equation 
(P$_{\rm VI}$), and developed in diverse directions 
in the next decade.  R. Fuchs considered a second order 
scalar ODE 
\[
    \frac{d^2 y}{d\lambda^2} + p(\lambda)y = 0 
\]
of Fuchsian type with four regular singularities, 
one of which is an apparent singularity. Garnier 
\cite{bib:Garnier1912} 
generalized the work of R. Fuchs in two different 
forms. One generalization is to consider more than 
four regular singularities. This leasd to a 
multi-variable generalization of P$_{\rm VI}$. 
The other is to include irregular singularities. 
The other five Painlev\'e equations
(P$_{\rm I}$ -- P$_{\rm V}$) can be derived 
from this generalization. Schlesinger 
\cite{bib:Schlesinger} obtained the so called 
Schlesinger equation from isomonodromic deformations 
of first order matrix ODE  
\[
    \frac{dY}{d\lambda} = M(\lambda) Y, 
\]
where $M(\lambda)$ is an $r \times r$ matrix of the form 
\[
    M(\lambda) = \sum_{i=1}^N \dfrac{A_i}{\lambda - t_i}. 
\]
Garnier noticed later \cite{bib:Garnier1926} that 
the $2 \times 2$ Schlesinger equation is equivalent 
to his isomonodromic deformations of 2nd order scalar 
Fuchsian equation.  

It is also in this decade that a link with 
``isospectral deformations'' was uncovered. 
This is again due to Garnier \cite{bib:Garnier1917}. 
He proposed an autonomous analogue of the Schlesinger 
equations, and pointed out that it can be integrated by 
Abelian functions. Remarkably, Garnier substantially 
arrived at the notion of isospectral deformations 
therein. Let us briefly review Garnier's discovery. 
Schlesinger's equation can be written 
\[
    \dfrac{\rd A_i}{\rd t_j} 
    = (1 - \delta_{ij}) \dfrac{[A_i,A_j]}{t_i - t_j} 
    - \delta_{ij} \sum_{k (\not= i)} \dfrac{[A_i,A_k]}{t_i - t_k}. 
\] 
This is a non-autonomous system, because the right hand side 
contains the independent variables $t = \{t_i\}$ explicitly. 
Garnier's proposal was to replace these $t_i$'s by constants, 
$t_i \to c_i$. The outcome is an autonomous system of the form 
\[
    \dfrac{\rd A_i}{\rd t_j} 
    = (1 - \delta_{ij}) \dfrac{[A_i,A_j]}{c_i - c_j} 
      - \delta_{ij} \sum_{k (\not= i)} \dfrac{[A_i,A_k]}{c_i - c_k}. 
\] 
As Garnier noticed, this gives isospectral deformations of 
$M(\lambda)$, namely, the characteristic polynomial 
$\det\Bigl( M(\lambda) - \mu I\Bigr)$ is invariant under 
the $t$ flows. The algebraic curve defined by the 
characteristic equation 
\[
    \det\Bigl( M(\lambda) - \mu I \Bigr) = 0 
\]
in the $(\lambda,\mu)$ plane (and its appropriate 
compactification) is nowadays called the ``spectral curve.''  
What Garnier did is to solve the above autonomous system 
in terms of Abelian functions on this algebraic curve. 

Isospectral problems of the same type were studied 
by Moser \cite{bib:Moser} in the 70's.  Moser proposed 
a matrix nonlinear system as a unified framework for 
a number of classically well known completely integrable 
dynamicals systems, such as the Neumann and Rosochatius 
systems, geodesic flows on an ellipsoid, etc.  Moser's 
idea and Munford's related work \cite{bib:Munford} 
were reformulated  by the Montreal group (Adams, Harnad, 
Hurtubise and Previato) \cite{bib:AHHP,bib:AHH} 
and Beauville \cite{bib:Beauville} to the isospectral 
problem of a rational matrix of the form $M(\lambda)$ (or 
$U + M(\lambda)$, where $U$ is a constant matrix). 
Harnad and his collaborators later applied their method 
to isomonodromic problems 
\cite{bib:Harnad-dual,bib:HTW,bib:HR,bib:HW}. 

Garnier's proposal, however, originally aimed at 
a quite different issue. He considered the autonomous 
system as a tool for studying asymptotic behavior of 
solutions of the Schlesinger equation in a neighborhood 
of singularities. A similar problem concerning the 
Painlev\'e equations  had been pursued by Boutroux 
\cite{bib:Boutroux}. Boutroux obtained an asymptotic 
expression of Painlev\'e transcendents as a ``modulated'' 
elliptic function. Here ``modulation'' means that 
parameters of the elliptic function also depend 
(but ``slowly'') on the independent variable.  
In Garnier's program, Boutroux's elliptic curve 
is replaced by a more general algebraic curve. 
Flaschka and Newell \cite{bib:FN} 
revisited Garnier's program in their study of 
isomonodromic and isospectral problems, and noted 
an important remark: They pointed out that a JWKB 
approximation converts the monodromy problem into 
a spectral problem. 

We reconsider this issue in the light of researches 
in the last ten years.  Since the end of the 80's, 
low dimensional string and topological field theories 
have provided new subjects of isomonodromic problems. 
A central subject is the so called ``string quations'' 
of two-dimensional quantum gravity \cite{bib:Moore-review}. 
They are the first (or second) Painlev\'e equation and 
its higher dimensional generalizations. These string 
equations have been studied from several different 
points of view. Among them, we are particularly interested 
in the approach \cite{bib:Novikov,bib:Krichever-ETH,
bib:Fucito-etal,bib:Vereschagin-string} 
from the ``Whitham averaging method'' \cite{bib:Whitham} 
(also referred to as the ``nonlinear JWKB method'' 
\cite{bib:D-Maslov}). This method may be viewed as 
a modernized version of Boutroux's analysis. We now 
consider the Schlesinger equation in the same philosophy. 

The Whitham averaging method, or the nonlinear JWKB method, 
covers a wide area of modulational phenomena in nonlinear 
waves. The most relevant for our problem is the case of 
modulation of Abelian function solutions (which are usually 
called ``finite-gap'' or ``quasi-periodic'' solutions 
\cite{bib:DMN,bib:Krichever1977}) to a soliton equation. 
An ``unmodulated'' finite-band solution generally takes 
the form 
\[
    u_0 = u_0 \Bigl( \sum_i U_i t_i \mid \{ I_n \} \Bigr), 
\]
where $u_0$ is an Abelian function, $U_i$'s are $g$-dimensional 
constant vectors ($g$ is the genus of the spectral curve), 
and $I_n$'s are other parameters of the solution. 
The parameters $U_i$'s and $I_n$'s are eventually determined 
by the spectral curve (as period integrals of meromorphic 
differentials).  If the problem in question contains a small 
parameter $\epsilon$, one may consider a solution with the 
following asymptotic form 
\[
    u \sim u_0 \Bigl( \sum_i U_i(T) t_i \mid \{I_n\} \Bigr) 
\]
as $\epsilon \to \infty$. The parameter $U_i$ and $I_n$ now 
depend on the ``slow variables'' 
\[
    T = \{ T_i \}, \quad T_i = \epsilon t_i. 
\]
This is the ``modulation'' of a finite-gap solution. 
The Whitham averaging method is a method to determine this 
slow dynamics in $T_i$ in the form of differential equations. 
This kind of differential equations are generally called 
``modulation equations'' (or ``Whitham equations'').  
In the case of finite-gap solutions, the modulation 
equation can be formulated as a dynamical system on 
the moduli space of spectral curves. 

These modulation equations of finite-gap solutions 
are known to possess a number of remarkable properties 
\cite{bib:DN-review}. It is Flaschka, Forest and McLaughlin 
\cite{bib:FFM} who first pointed out that this type of 
modulation equations have a universal structure. They 
demonstrated, in the case of the KdV equation, that the 
modulation equation boils down to the universal form 
\[
    \frac{\rd}{\rd T_i} d\Omega_j 
    = \frac{\rd}{\rd T_j} d\Omega_i, 
\]
where $d\Omega_i$'s are meromorphic differentials on 
the spectral curve.  Krichever \cite{bib:Krichever-Whitham} 
and Dubrovin \cite{bib:Dubrovin-Whitham} presented 
an abstract reformulation of this type of equations 
(``Whitham hierarchies''), and constructed many special 
solutions (with applications to geometry and physics). 

Our concern lies in special solutions of the universal 
Whitham hierarchy that represent slow dynamics of a 
spectral curve in isomonodromic problems.  Remarkably, 
it seems likely that this class of solutions of the 
universal Whitham hierarchy are always characterized 
by a differential equation of the form 
\[
    \frac{\rd}{\rd T_i}dS = d\Omega_i, \quad 
    dS = \mu d\lambda. 
\]
This is indeed the case for the string equations 
\cite{bib:Novikov,bib:Krichever-ETH,
bib:Fucito-etal,bib:Vereschagin-string}. 
In this paper, we shall derive a modulation equation 
of this form from the Schlesinger equation.  

We use a very heuristic method to derive the modulation 
equation.  This heuristic method was developed in an 
attempt \cite{bib:TN} at an isomonodromic interpretation 
of integrable structures in supersymmetric gauge theories 
\cite{bib:SW-integrable}.  
The modulation equation turns out to possess almost the 
same properties as the so called ``Seiberg-Witten solutions'' 
of $N=2$ supersymmetric gauge theories.  In particular, we 
introduce a period map of Seiberg-Witten map, and prove that 
the inverse period map solves the modulation equation. This 
also reveals the existence of another set of commuting flows 
on the moduli space of spectral curves.  We also show that 
the notion of prepotential can be generalized to this case. 
These results will be strong evidence for the validity of 
the heuristic derivation. 

This paper is organized as follows. Sections 2 and 3 are of 
preliminary nature. In Section 2, we review basic properties 
of the Schlesinger equation.  In Section 3, we consider the 
geometric structure of spectral curves along the lines of 
approach by the Montreal group and Beauville. Section 4, 5, 
and 6 are focussed on the derivation of the modulation equation. 
We begin with a reformulation of the Schlesinger equation 
in Section 4. The reformulated Schlesinger equation has a 
small parameter $\epsilon$.  Garnier's autonomous system 
emerges in the limit of $\epsilon \to 0$.  Our modulation 
equation is derived in Section 5, along with comments on other 
possible approaches. The structure of meromorphic differentials 
$d\Omega_i$, which are also basic constituents of our 
modulation equation, is specified in Section 6. Section 7 
is devoted to solving the modulation equation by the inverse 
period map. Section 8 deals with the notion of prepotential.  
We conclude this paper in Section 9.

\section{Schlesinger Equation}

In this section, we review basic properties of the Schlesinger 
equation. For details and related topics, we refer to a series 
of papers by the Kyoto school (Jimbo, Miwa, M\^ori, Sato and Ueno) 
\cite{bib:JMMS,bib:JMU,bib:JM}. 

\subsection{Coadjoint Orbit and Hamiltonian Structure} 

Let $\gl(r,\bfC)^N$ denote a direct sum of $N$ copies of 
$\gl(r,\bfC)$. This is the space of $N$-tuples 
$(A_1,\cdots,A_N)$ of $r \times r$ matrices.  $\GL(r,\bfC)$ 
acts on this space by the diagonal coadjoint action: 
$A_i \mapsto g A_i g^{-1}$. The Schlesinger equation 
can be written 
\begin{equation}
    \dfrac{\rd A_i}{\rd t_j} 
      = \left[ A_i,  
          (1 - \delta_{ij}) \dfrac{A_j}{t_i - t_j} 
          - \delta_{ij} \sum_{k(\not= i)} 
             \dfrac{A_k}{t_i - t_k} \right], 
\end{equation}
each coadjoint orbit $\calO_i$ is left invariant under the 
$t$-flows. Thus the Schlesinger equation is actually a 
collection of non-autonomous dynamical systems on a 
direct product  $\calO_1 \times \cdots \times \calO_N$ 
of coadjoint orbits in $\gl(r,\bfC)$. Usually, only semi-simple 
orbits are considered; such an orbit is labeled by the 
eigenvalues $\theta_{i\alpha}$ ($\alpha = 1, \cdots, r$) 
of $A_i$. In other words, these eigenvalues (and, in general, 
the Jordan canonical form of $A_i$'s) are invariants of the 
Schlesinger equation. 

Actually, there are some extra invariants. They are the 
matrix elements of 
\begin{equation}
    A_\infty = - \sum_{i=1}^N A_i, 
\end{equation}
which are invariant under the Schlesinger equation: 
\begin{equation}
    \dfrac{\rd A_\infty}{\rd t_i} = 0.
\end{equation}
This matrix, too, is usually assumed to be semi-simple, 
and it is customary to diagonalize this matrix in 
advance by a constant ``gauge transformation'' 
$A_i \mapsto C A_i C^{-1}$. Thus only the eigenvalues 
$\theta_{\infty \alpha}$ ($\alpha = 1,\cdots,r$) of 
$A_\infty$ are nontrivial invariants.

Geometrically, this gauge-fixing may be interpreted as 
the Marsden-Weinstein construction of a ``reduced phase 
space.'' The ``unreduced phase space'' is a coadjoint orbit 
$\calO_1 \times \cdots \calO_N \times \calO_\infty$ in 
the vector space $\gl(r,\bfC)^{N+1}$ of $(N+1)$-tuples 
$(A_1,\cdots,A_N,A_\infty)$.  In order to reproduce the 
Schlesinger equation, one has to impose the linear constraint 
\begin{equation}
    \sum_{i=1}^N A_i + A_\infty = 0, 
\end{equation}
and ``gauge away'' redundant degrees of freedom by the 
action of $\GL(r,\bfC)$. The left hand side of the linear 
constraint is essentially a moment map of this diagonal 
$\GL(r,\bfC)$ action.

Analytically, as we see show below, the coadjoint orbit 
invariants $\theta_{i\alpha}$ ($i = 1,\cdots,N,\infty$, 
$\alpha = 1,\cdots,r$) give local monodromy exponents of 
Schlesinger's monodromy problem.

The coadjoint orbit structure leads to a Hamiltonian 
formalism of the Schlesinger equation.  Let us introduce 
a Poisson structure on the vector space $\gl(r,\bfC)^N$ 
by defining the Poisson bracket of matrix elements of 
$A_i = (A_{i,\alpha\beta})$ as: 
\begin{equation}
    \{ A_{i,\alpha\beta}, A_{j,\rho\sigma} \} 
    = \delta_{ij} \Bigl( 
       - \delta_{\beta\rho} A_{i,\alpha\sigma} 
       + \delta_{\sigma\alpha} A_{i,\rho\beta} \Bigr). 
\end{equation}
In each component of the direct sum, this is just the 
ordinary Kostant-Kirillov Poisson bracket. The Schlesinger 
equation can be written in the Hamiltonian form
\begin{equation}
    \dfrac{\rd A_j}{\rd t_i} = \{ A_j, H_i \}, 
\end{equation}
where the Hamiltonians are given by 
\begin{equation}
    H_i = \Res_{\lambda=T_i} \frac{1}{2} \Tr M(\lambda)^2 
    = \sum_{j(\not= i)} \Tr\left( 
            \dfrac{A_i A_j}{t_i - t_j} \right), 
\end{equation}
and involutive, 
\begin{equation}
    \{ H_i, H_j \} = 0. 
\end{equation}

\subsection{Isomonodromic Deformations}

The Schlesinger equation gives isomonodromic deformations of 
the first order ODE 
\begin{equation}
    \frac{dY}{d\lambda} = M(\lambda)Y 
\end{equation}
with the rational coefficient matrix 
\begin{equation}
    M(\lambda) = \sum_{i=1}^N \dfrac{A_i}{\lambda - t_i}. 
\end{equation}
Note that $A_\infty$ is the residue of $M(\lambda)$ at  
$\lambda = \infty$.

Usually, this type of isomonodromic problems are considered 
under the following 
\medskip

\noindent {\bf Assumption} 
\begin{itemize}
\item The residue matrices $A_i$ ($i = 1,\cdots,N,\infty$) 
      are diagonalizable. 
\item The eigenvalues $\theta_{i\alpha}$ ($\alpha = 1,\cdots,r$) 
      of each residue matrix $A_i$ have no integer difference, 
      i.e., $\theta_{i\alpha} - \theta_{i\beta} \not\in \bfZ$ if 
      $\alpha \not= \beta$. 
\end{itemize} 
We assume them throughout this paper.  These assumptions 
ensure that local solutions at the singular points 
$\lambda = t_1,\cdots,t_N,\infty$ develop no logarithmic 
term (see below).

The isomonodromic deformations are generated by the deformation 
equations 
\begin{equation}
    \frac{\rd Y}{\rd t_i} = - \dfrac{A_i}{\lambda - t_i} Y. 
\end{equation}
These deformation equations and the above first order ODE 
comprise an ``auxiliary linear problem'' of the Schlesinger 
equation.  Its Frobenius integrability conditions can be written 
in the ``zero-curvature form'' 
\begin{equation}
    \left[ \dfrac{\rd}{\rd t_j} + \dfrac{A_i}{\lambda - t_i}, \ 
       M(\lambda) - \dfrac{\rd}{\rd \lambda} \right] = 0,  
    \quad 
    \left[ \dfrac{\rd}{\rd t_i} + \dfrac{A_i}{\lambda - t_i}, \ 
       \dfrac{\rd}{\rd t_j} + \dfrac{A_j}{\lambda - t_j} \right] = 0, 
\end{equation}
and one can easily check that these zero-curvature equations 
are equivalent to the Schlesinger equation.

\subsection{Local Solutions at Singular Points and Tau Function}

Since $\lambda = t_1,\cdots,t_N$ and $\lambda=\infty$ are 
regular singularities of the above first order ODE, one can 
construct a local solution of the following form at each of 
these singular points: 
\begin{itemize}
  \item Local solution at $\lambda = t_i$: 
  \begin{equation}
    Y_i = \Yhat_i \cdot (\lambda - t_i)^{\Theta_i}, \quad 
    \Yhat_i = \sum_{n=0}^\infty Y_{in} (\lambda - t_i)^n. 
  \end{equation}
\item Local solution at $\lambda = \infty$: 
  \begin{equation}
    Y_\infty = \Yhat_\infty \cdot \lambda^{-\Theta_\infty}, \quad 
    \Yhat_\infty = \sum_{n=0}^\infty Y_{\infty n} \lambda^{-n}. 
  \end{equation}
\end{itemize}
Here $Y_{in}$ are $r \times r$ matrices, the leading coefficients 
$Y_{i0}$ and $Y_{\infty 0}$ are invertible, and $\Theta_i$ and 
$\Theta_\infty$ are diagonal matrices of local monodromy exponents. 
Inserting these expressions into the first order ODE gives the 
relations
\begin{equation}
    A_i= Y_{i0} \Theta_i Y_{i0}^{-1}, \quad 
    i = 1,\cdots,N,\infty. 
\end{equation}
In particular, local monodromy exponents coincide with the 
eigenvalues of $A_i$: 
\begin{equation}
    \Theta_i = \diag( \theta_{i1}, \cdots, \theta_{ir} ). 
\end{equation}
It is not hard to check that these expressions of $A_i$ and 
$\Theta_i$ are also consistent with the other equations of 
the auxiliary linear problem.

The $\tau$ function of the Schlesinger equation is defined 
in two equivalent ways. One way is to define $\log\tau$ as a 
potential of the Hamiltonians $H_i$:
\begin{equation}
    d\log\tau = \sum_{i=1}^N H_i dt_i. 
\end{equation}
Another equivalent definition, which is more suited for 
generalization, is based on the equations 
\begin{equation}
    \frac{\rd \log \tau}{\rd t_i} = \Tr \Theta_i Y_{i0}^{-1}Y_{i1}. 
\end{equation}
The equivalence can be verified as follows: 
\begin{eqnarray*}
    H_i &=& \sum_{j(\not= i)} \Tr \dfrac{A_i A_j}{t_i - t_j}  \\
    &=& \Res_{\lambda=t_i} 
          \Tr \dfrac{A_i}{\lambda - t_i} M(\lambda) \\
    &=& \Res_{\lambda=t_i} 
          \Tr \dfrac{Y_{i0} \Theta_i Y_{i0}^{-1}}{\lambda - t_i} 
              \frac{\rd Y_i}{\rd \lambda} Y_i^{-1}  \\
    &=& \Res_{\lambda=t_i}
          \Tr \dfrac{\Theta_i}{\lambda - t_i}
              Y_{i0}^{-1} \frac{\rd Y_i}{\rd \lambda} 
              Y_i^{-1} Y_{i0} \\
    &=& \Tr \Theta_i Y_{i0}^{-1} Y_{i1}. 
\end{eqnarray*}
The closedness of the 1-form $\sum H_i dt_i$, or equivalently 
the integrability condition 
\begin{equation}
    \dfrac{\rd H_i}{\rd t_j} = \dfrac{\rd H_j}{\rd t_i}, 
\end{equation}
is ensured by the Schlesinger equation itself.

\section{Spectral Curve}

By ``spectral curve,'' we mean the plane algebraic 
curve defined on the $(\lambda,\mu)$ plane by 
\begin{equation}
    \det\Bigl( M(\lambda) - \mu I \Bigr) = 0 
\end{equation}
and its suitable compactification.  We first discuss 
its roles in isomonodromic and isospectral problems, 
then consider its geometric properties.

\subsection{Spectral Curve in Isomonodromic Problem}

Isomonodromic deformations such as the Schlesinger 
equations are non-isospectral, namely, the spectral 
curve varies in deformations. 

Let us present an interesting formula (essentially 
due to Vereschagin \cite{bib:Vereschagin-painleve}) 
which show {\it qualitatively} that the characteristic 
polynomial of $M(\lambda)$ varies under isomonodromic 
deformations.  First, by the the well known identities 
of linear algebra, we have 
\begin{eqnarray*} 
    \frac{\rd}{\rd t_i} \log \det \Bigl( M(\lambda) - \mu I \Bigr) 
    &=& \frac{\rd}{\rd t_i} \Tr\log 
              \Bigl( M(\lambda) - \mu I \Bigr)  \\
    &=& \Tr \frac{\rd M(\lambda)}{\rd t_i} 
              \Bigl( M(\lambda) - \mu I \Bigr)^{-1}. 
\end{eqnarray*}
By the first equation of the zero-curvature representation, 
$\rd M(\lambda)/\rd t_i$ can be rewritten 
\[
    \frac{\rd M(\lambda)}{\rd t_i} 
    = \dfrac{A_i}{(\lambda - t_i)^2} 
      - \left[ \dfrac{A_i}{\lambda - t_i}, M(\lambda) - \mu I \right]. 
\]
The second term on the right hand side has no contribution, 
because of the identity 
\[
    \Tr \left[ \dfrac{A_i}{\lambda - t_i}, M(\lambda) - \mu I \right] 
        \Bigl( M(\lambda) - \mu \Bigr)^{-1} = 0. 
\]
Thus, eventually, we obtain the formula 
\begin{equation}
    \frac{\rd}{\rd t_i} \log\det \Bigl( M(\lambda) - \mu I \Bigr) 
    = \Tr \dfrac{A_i}{(\lambda - t_i)^2} 
          \Bigl( M(\lambda) - \mu I \Bigr)^{-1}. 
\end{equation}
This clearly shows that the $t$-dependence of the characteristic 
polynomial of $M(\lambda)$ is driven by the ``anomalous'' term 
$A_i/(\lambda - t_i)^2$.

\subsection{Spectral Curve in Isospectral Problem} 

We now turn to Garnier's autonomous analogue of the 
Schlesinger equation.  This equation has the following 
zero-curvature representation: 
\begin{equation}
    \left[ \dfrac{\rd}{\rd t_i} + \dfrac{A_i}{\lambda - c_i}, \ 
      M(\lambda) \right] = 0, 
    \quad 
    \left[ \dfrac{\rd}{\rd t_i} + \dfrac{A_i}{\lambda - c_i}, \ 
      \dfrac{\rd}{\rd t_j} + \dfrac{A_j}{\lambda - c_j} \right] = 0. 
\end{equation}
Repeating the same calculations as above, we now find that 
\begin{equation}
    \frac{\rd}{\rd t_i} \det\Bigl( M(\lambda) - \mu I \Bigr) = 0, 
\end{equation}
because there is no ``anomalous'' term like $A_i/(\lambda - t_i)^2$. 
Thus one can confirm that Garnier's autonomous system is 
indeed an isospectral problem. An auxiliary linear problem 
is given by 
\begin{equation} 
    \mu \psi = M(\lambda) \psi, \quad 
    \frac{\rd \psi}{\rd t_i} = - \dfrac{A_i}{\lambda - c_i} \psi. 
\end{equation} 
Here $\psi$ is understood to be a column vector. The first 
equation means that $\psi$ is an eigenvector of $M(\lambda)$ 
with eigenvalue $\mu$. The second set of equations generate 
isospectral deformations.

It is nowadays well known that this type of isospectral 
problems can be mapped to linear flows on the Jacobian 
variety of the spectral curve \cite{bib:DMN,bib:Krichever1977}. 
Reproducing the original nonlinear problem is identical 
to Jacobi's inversion problem, and indeed solvable by 
theta functions (or Baker-Akhiezer functions). The solution 
thus obtained is written in terms of period integrals and 
Abelian functions.

Although isomonodromic problems (in a generic case) cannot 
be solved in that way, the notion of spectral curve still 
plays a role in the study of Hamiltonian structures.  
Indeed, Harnad and Wisse \cite{bib:Harnad-dual,bib:HW} 
presented a construction of special Darboux coordinates 
(``spectral Darboux coordinates'') using the language 
of spectral curves.  The case of $r = 2$ is particularly 
interesting, because this is the case where the 6th 
Painlev\'e equation (P$_{\rm VI}$) and Garnier's 
multi-variable version of P$_{\rm VI}$ (called the 
``Garnier system'' by Okamoto \cite{bib:Okamoto}) emerge. 
The spectral Darboux coordinates in this case coincide 
with Okamoto's Darboux coordinates for the ``Garnier system,'' 
which Okamoto discovered without using the notion of spectral 
curve.

\subsection{Geometry of Spectral Curves} 

The structure of spectral curves of the above type 
has been elucidated in detail by the Montreal group 
\cite{bib:AHHP,bib:AHH} and Beauville \cite{bib:Beauville}. 
We present basic part of their results, which will be used 
in the subsequent sections. In the following, the poles of 
$M(\lambda)$ are written $c_i$ rather than $T_i$.

\subsubsection{Spectral Curve on Plane.} 

Let $C_0$ be the spectral curve on the $(\lambda,\mu)$ plane: 
\begin{equation} 
    F(\lambda,\mu) = \det\Bigl( M(\lambda) - \mu I \Bigr) = 0. 
\end{equation}
This becomes a ramified covering of the punctured Riemann sphere, 
\begin{equation}
    \pi: C_0 \to \bfC P^1 \setminus \{ c_1,\cdots,c_N,\infty \}, 
    \quad  \pi(\lambda,\mu) = \lambda. 
\end{equation}
For a generic value of $\lambda$, the inverse image 
$\pi^{-1}(\lambda)$ consists of $r$ points 
$(\lambda,\mu_\alpha)$ ($\alpha = 1,\cdots,r$). The 
$\mu$-coordinates $\mu_\alpha$ of these points are 
eigenvalues of $M(\lambda)$. In a neighborhood of 
$\lambda = c_i$, $\mu_\alpha$'s behave as 
\begin{equation}
    \mu_\alpha = \dfrac{\theta_{i\alpha}}{\lambda - c_i} 
       + \mbox{non-singular}, 
\end{equation}
where $\theta_{i\alpha}$ are the eigenvalues of $A_i$. 
Similarly, in a neighborhood of $\lambda = \infty$, 
\begin{equation}
    \mu_\alpha = - \theta_{\infty\alpha} \lambda^{-1} 
    + O(\lambda^{-2}), 
\end{equation}
where $\theta_{\infty\alpha}$ are the eigenvalues of 
$A_\infty = - \sum_{i=1}^N A_i$. 
(Of course, the numbering of $\mu_\alpha$'s is meaningful 
only locally.)

\subsubsection{Compactification of Spectral Curve} 

We now compactify $C_0$ by adding several points over 
the punctures of the Riemann sphere.  In a neighborhood 
of $\lambda = c_i$, let us consider the following 
$\mutilde$ in place of $\mu$: 
\begin{equation}
    \mutilde = f(\lambda) \mu, \quad 
    f(\lambda) = \prod_{i=1}^N (\lambda - c_i). 
\end{equation}
In terms of the new coordinates $(\lambda,\mutilde)$, 
the equation of the spectral curve becomes 
\begin{equation}
    \Ftilde(\lambda,\mutilde) 
    = \det \Bigl( f(\lambda)M(\lambda)  - \mutilde I \Bigr) = 0. 
\end{equation}
The inverse image $\pi^{-1}(\lambda)$ consists of $r$ points 
$(\lambda,\mutilde_\alpha)$  ($r = 1,\cdots,r)$ such that 
\begin{equation}
    \mutilde_\alpha = 
      f'(c_i) \theta_{i\alpha} + O(\lambda- c_i) 
\end{equation}
as $\lambda \to c_i$. Since the eigenvalues 
$\theta_{i\alpha}$ ($\alpha = 1,\cdots,r$) 
are pairwise distinct (recall the assumptions in Section 2), 
we add to $C_0$ extra $r$ points $(\lambda,\mutilde) 
= \bigl(c_i,f'(c_i)\theta_{i\alpha} \bigr)$ to fill the holes 
above $\lambda = c_i$. 

Similarly, in a neighborhood of $\lambda = \infty$, we use 
\begin{equation}
    \mutilde = \lambda \mu 
\end{equation}
in place of $\mu$.  $\pi^{-1}(\lambda)$ now consists of the 
$r$ points $(\lambda,\mutilde_\alpha)$ ($r = 1,\dots,r)$ 
such that 
\begin{equation}
    \mutilde_\alpha = - \theta_{\infty\alpha} 
      + O(\lambda^{-1})
\end{equation}
as $\lambda \to \infty$. The eigenvalues 
$\theta_{\infty\alpha}$ ($r = 1,\cdots,r$), too, are 
pairwise distinct. Therefore we add to $C_0$ the $r$ points 
$(\lambda,\mutilde) = (\infty,-\theta_{\infty\alpha})$ 
to fill the holes above $\lambda = \infty$. 

Thus, by adding altogether $rN + r$ points to $C_0$, we obtain 
a compactification $C$ of $C_0$.  The covering map  $\pi$ 
uniquely extends to $C$, and gives a ramified covering 
$\pi: C \to \bfC P^1$ of the Riemann sphere.

\subsubsection{Genus of Compactified Spectral Curve}

If there are multiple eigenvalues of $A_i$, the compactified 
spectral curve $C$ has singular points (nodes) over 
$\lambda = c_i$.  In that case, one has to take a 
desingularization $\rho: \Ctilde \to C$ 
for further consideration.  Fortunately, this does not occur 
in our case because of the assumptions introduced in Section 2. 
One can show, by a standard method, that $C$ has the genus
\begin{equation}
    g = \frac{1}{2}(r-1)(rN - r - 2). 
\end{equation}

\subsubsection{Structure of Characteristic Polynomial}

In order to examine the structure of the characteristic 
polynomial of $M(\lambda)$ (or, rather, $f(\lambda)M(\lambda)$), 
we take the rational matrix 
\begin{equation}
    M^0(\lambda) = \sum_{i=1}^N \dfrac{A^0_i}{\lambda - c_i}, 
    \quad 
    A^0_i = \diag( \theta_{i1}, \cdots, \theta_{ir} ), 
\end{equation}
as a reference point on the coadjoint orbit that $M(\lambda)$ 
belongs to. Now compare the  characteristic polynomials 
$\Ftilde(\lambda,\mutilde)$ and $\Ftilde^0(\lambda,\mutilde)$ 
of $f(\lambda)M(\lambda)$ and $f(\lambda)M^0(\lambda)$. 
Note, first, that 
$\Ftilde(\lambda,\mutilde) - \Ftilde^0(\lambda,\mu)$ 
vanishes at $\lambda = c_i$. This will be obvious from 
the following expression of these polynomials as 
$\lambda \to c_i$: 
\begin{eqnarray*}
    \Ftilde(\lambda,\mutilde) = 
      \det\Bigl( f'(c_i)A_i - \mutilde I \Bigr) + O(\lambda - c_i), \\
    \Ftilde^0(\lambda,\mutilde) = 
      \det\Bigl( f'(c_i)A^0_i - \mutilde I \Bigr) + O(\lambda - c_i). 
\end{eqnarray*}
(The two determinants on the right hand side are equal, 
because $A_i$ and $A^0_i$ are on the same coadjoint orbit.)  
Therefore $\Ftilde(\lambda,\mutilde) - \Ftilde^0(\lambda,\mu)$ 
is divisible by $f(\lambda)$. Furthermore, these 
characteristic polynomials have the following expansion 
in powers of $\mutilde$: 
\begin{eqnarray*}
    \Ftilde(\lambda,\mutilde) = 
      (-\mutilde)^r + \Tr M(\lambda) (-\mutilde)^{r-1} + \cdots, \\
    \Ftilde^0(\lambda,\mutilde) = 
      (-\mutilde)^r + \Tr M^0(\lambda) (-\mutilde)^{r-1} + \cdots. 
\end{eqnarray*}
Since the first two terms on the right hand side are equal, 
respectively, the difference of the two polynomials contains 
no terms proportional to $\mu^r$ and $\mu^{r-1}$.  From these 
observations, one can conclude that the difference of the 
characteristic functions can be written 
\[ 
    \Ftilde(\lambda,\mutilde) - \Ftilde(\lambda,\mutilde) = 
      f(\lambda) \sum_{\ell=2}^r p_\ell(\lambda) \mutilde^{r-\ell}, 
\] 
where $p_\ell(\lambda)$'s are polynomial functions of $\lambda$. 
A simple power counting argument (assigning weight $1$ to 
$\lambda$ and weight $N-1$ to $\mutilde$) shows that the degree 
of $p_\ell(\lambda)$ does not exceed the positive integer 
\begin{equation}
    \delta_\ell = (N-1)\ell - N. 
\end{equation}
Therefore $p_\ell(\lambda)$ can be written 
\[
    p_\ell(\lambda) = 
      \sum_{m=0}^{\delta_\ell} h_{m\ell} \lambda^m. 
\]
The leading coefficient of this polynomial is a 
function of $\theta_{i\alpha}$'s only, because it 
can be writen 
\begin{equation}
    h_{\delta_\ell,\ell} 
    = (-1)^r \Bigl( \sigma_\ell(A_\infty) 
                   -\sigma_\ell(A^0_\infty) \Bigr), 
\end{equation}
where $\sigma_\ell(A_\infty)$ and $\sigma_\ell(A^0_\infty)$ 
are the $\ell$-th elementary symmetric function of 
$A_\infty$ and $A^0_\infty = - \sum_{i=1}^r A^0_i$. 
Thus, eventually, 
we arrive at the following expression of 
$\Ftilde(\lambda,\mutilde)$:
\begin{equation}
    \Ftilde(\lambda,\mutilde) = 
      \Ftilde^0(\lambda,\mutilde) 
      + \sum_{\ell=2}^r \sum_{m=0}^{\delta_\ell} 
          h_{m\ell} \lambda^m \mutilde^{r-\ell}. 
\end{equation}

\subsubsection{Parameters of Spectral Curve} 

Apart from the leading coefficients, the coefficients 
$h_{m\ell}$ ($\ell = 2,\cdots,r$, $m = 0,\cdots,\delta_\ell - 1$) 
of $p_\ell(\lambda)$'s give arbitrary parameters (``moduli'') 
of the spectral curve. Their total number coincides with 
the genus of the spectral curve: 
\begin{equation}
    \sum_{\ell=2}^r \delta_\ell 
    = \sum_{\ell=2}^r \Bigl( (N-1)\ell - N \Bigr) 
    = g. 
\end{equation}
In the isospectral problem, these parameters $h_{m\ell}$ 
are constants of motion (Hamiltonians of commuting 
isospectral flows). They should not be confused with the 
Hamiltonians $H_i$ of the isomonodromic deformations. 

Actually, the characteristic polynomial has yet another 
set of parameters --- the position $c_i$ of poles of 
$M(\lambda)$.  They play the role of deformation 
variables in isomonodromic problem. 

In summary, the spectral curve has three distinct sets of 
parameters: 
\begin{itemize}
\item Position of poles $c_i$ ($i=1,\cdots,N$). 
\item Coadjoint orbit invariants $\theta_{i\alpha}$ 
      ($i=1,\cdots,N,\infty$). 
\item Isospectral invariants $h_{m\ell}$ 
      ($\ell=2,\cdots,r$, $m=0,\ldots,\delta_\ell-1$). 
\end{itemize}

\section{Schlesinger Equation with Small Parameter}

\subsection{Reformulation Including Small Parameter}

The first step towards the derivation of our modulation 
equation is to reformulate the Schlesinger equation by the 
following substitution rule:
\begin{equation}
    \frac{\rd}{\rd t_i} \to \epsilon \frac{\rd}{\rd T_i}, 
    \quad 
    \frac{\rd}{\rd \lambda} \to \epsilon \frac{\rd}{\rd \lambda}, 
    \quad 
    \dfrac{A_i}{\lambda - t_i} \to \dfrac{A_i}{\lambda - T_i}. 
\end{equation} 
Note, in particular, that the deformation variables are 
renamed as $t_i \to T_i$.  $T_i$'s will play the role of 
``slow variables.''  The reformulated Schlesinger equation 
reads: 
\begin{equation}
    \epsilon \dfrac{\rd A_i}{\rd T_j} 
    = (1 - \delta_{ij}) \dfrac{[A_i,A_j]}{T_i - T_j} 
      -  \delta_{ij} \sum_{k (\not= i)} \dfrac{[A_i,A_k]}{T_i - T_k}. 
\end{equation}
An auxiliary linear problem is given by 
\begin{equation}
    \epsilon \frac{\rd Y}{\rd \lambda} = M(\lambda) Y, \quad 
    \epsilon \frac{\rd Y}{\rd T_i} = - \dfrac{A_i}{\lambda - T_i} Y. 
\end{equation}
The above $\epsilon$-dependent Schlesinger equation can be indeed 
reproduced from the the Frobenius integrability conditions 
\begin{eqnarray}
    \left[ \epsilon \dfrac{\rd}{\rd T_j} + \dfrac{A_i}{\lambda - T_i}, \ 
       M(\lambda) - \epsilon \dfrac{\rd}{\rd \lambda} \right] = 0,  
    \nonumber \\ 
    \left[ \epsilon \dfrac{\rd}{\rd T_i} + \dfrac{A_i}{\lambda -T_i}, \ 
       \epsilon \dfrac{\rd}{\rd T_j} + \dfrac{A_j}{\lambda - T_j} 
          \right] = 0. 
\end{eqnarray}

A few comments on the above reformulation will be in order: 

(i) Reformulating the Schlesinger equation as above is 
inspired by the work of Vereschagin 
\cite{bib:Vereschagin-painleve}, who considered all the six 
Painlev\'e equations (P$_{\rm I}$--P$_{\rm VI}$) in such an 
$\epsilon$-dependent form. This is different from the way 
Boutroux \cite{bib:Boutroux} and Garnier \cite{bib:Garnier1917} 
derived an Abelian function approximation, 
but, as Vereschagin stresses, they are asymptotically related. 

(ii) The string equations of two-dimensional quantum gravity, 
which are P$_{\rm I}$, P$_{\rm II}$ and their higher order 
generalization, contain such a small parameter from the 
beginning. The small parameter i interpreted as 
``string coupling constant'' \cite{bib:Moore-review}.

\subsection{Isospectral Deformations as First Approximation}

One can now see, at least intuitively, that an isospectral 
problem emerges from the above $\epsilon$-dependent 
Schlesinger equation.  Let us introduce the ``fast variables''
\begin{equation}
    t_i = \epsilon^{-1} T_i, 
\end{equation}
and rewrite the above equation as 
\begin{equation}
    \dfrac{\rd A_i}{\rd t_j} 
    = (1 - \delta_{ij}) \dfrac{[A_i,A_j]}{T_i - T_j} 
    - \delta_{ij} \sum_{k (\not= i)} \dfrac{[A_i,A_k]}{T_i - T_k}.  
\end{equation}
Suppose we now observe this system {\it in the scale of the 
fast variables\/} $t_i$. In this scale, $T_i$'s may be 
treated as being {\it approximately constant}, because 
a finite displacement in $t_i$'s corresponds to a small 
(i.e., $O(\epsilon)$) displacement in $T_i$'s. If 
$T_i$'s were true constants, the above equation would 
be exactly Garnier's autonomous system. Thus, observed 
in the scale of the fast variables, our $\epsilon$-dependent 
Schlesinger equation looks {\it approximately} like an 
isospectral problem. 

Of course, this is no more than an approximation.  
In fact, $T_i$'s are not constant, but vary slowly in 
the order of $O(\epsilon)$. Accordingly, the spectral 
curve, too, deforms slowly because the defining 
equation of the spectral curve contains $T_i$'s 
explicitly. A precise description of the isospectral 
approximation has to take into account such slow 
deformations of the spectral curve. 

It is, however, not only $T_i$'s that vary; the isospectral 
invariants $h_{m\ell}$ also change values slowly. They are 
both responsible for deformations of the spectral curve. 
In order to see how this occurs, recall the derivative 
formula of the characteristic polynomial under isomonodromic 
deformations (Section 3.1). In the present setting, 
this formula takes the form 
\begin{equation}
    \frac{\rd}{\rd T_i} \log\det \Bigl( M(\lambda) - \mu I \Bigr) 
    = \Tr \dfrac{A_i}{(\lambda - T_i)^2} 
          \Bigl( M(\lambda) - \mu I \Bigr)^{-1}. 
\end{equation}
The right hand side plays the role of a ``driving force'' 
for slow deformations of the characteristic polynomial. 
More precisely, in order to extract a true driving force 
of slow deformations, one has to take an average of the 
right hand side over the quasi-periodic motion in the 
fast variables $t_i$. This is a central idea of 
Vereschagin's averaging method 
\cite{bib:Vereschagin-string,bib:Vereschagin-painleve}. 

Meanwhile, the coadjoint orbit invariants $\theta_{i\alpha}$ 
remain constant, because they are also invariant of 
isomonodromic deformations.

\subsection{Concept of Multiscale Analysis}

Although frequently lacking mathematical rigor, 
``multiscale analysis'' is widely accepted in applied 
mathematics as a powerful tool for dealing with this 
kind of problems \cite{bib:multiscale}. 
The Whitham averaging method, too, has been developed 
in the framework of multiscale analysis. The nonlinear 
JWKB method of Dobrokhotov and Maslov \cite{bib:D-Maslov} 
is an attempt at a rigorous reformulation of multiscale 
analysis. Let us show how our problem may be formulated 
in the language of multiscale analysis. 

A key of multiscale analysis is to treat the fast 
variables $t = \{ t_i \}$ and the slow variables 
$T = \{ T_i \}$ as {\it independent variables\/}. 
$A_i$'s are thus assumed to be a function of $t$ and 
$T$ (and the small parameter $\epsilon$), 
\begin{equation}
    A_i = A_i(t,T,\epsilon). 
\end{equation}
The relation $t_i = \epsilon^{-1} T_i$ is imposed only 
in the differential equation in question. (For a more 
precise description, one may introduce a series of 
``slower'' variables $t^{(2)}$, $t^{(3)}$, $\cdots$, 
related with $t_k$'s as 
\begin{equation}
    t^{(k)}_i = \epsilon^k t_i. 
\end{equation}
For our purpose, only the first two scales are sufficient.) 

In this multiscale ansatz, derivative terms in the equation 
are given by a sum of $t$-derivatives and $T$-derivatives. 
In terms of differential operators, this amounts to substituting 
\begin{equation}
    \epsilon \frac{\rd}{\rd T_i} \to 
    \frac{\rd}{\rd t_i} + \epsilon \frac{\rd}{\rd T_i}. 
\end{equation}

We now assume an asymptotic expansion of the form 
\begin{equation}
    A_i(t,T,\epsilon) = 
      A^{(0)}_i(t,T) + A^{(1)}_i(t,T) \epsilon + \cdots, 
\end{equation}
and plug all these stuff into the Schlesinger equation. 
From each order of $\epsilon$, we obtain a differential 
equation for the coefficients of the above expansion.

The lowest order equation is give by 
\begin{equation}
    \dfrac{\rd A^{(0)}_i}{\rd t_j} 
    = (1 - \delta_{ij}) \dfrac{[A^{(0)}_i,A^{(0)}_j]}{T_i - T_j} 
    - \delta_{ij} \sum_{k (\not= i)} 
                        \dfrac{[A^{(0)}_i,A^{(0)}_k]}{T_i - T_k}. 
\end{equation}
If we consider $T_i = c_i$, this is nothing but Garnier's 
autonomous system. Note that this derivation of Garnier's 
isospectral problem is more understandable than the 
intuitive derivation in the last subsection.  In the derivation 
of the last subsection, $t$ and $T$ were not independent 
and constrained by the relation $t_i = \epsilon^{-1}T_i$, 
thereby we had to say that $T$ is ``approximately constant''; 
in the setting of multiscale analysis, $t$ and $T$ are 
{\it independent\/}. This shows a conceptual advantage of 
multiscale analysis. 

The lowest order equation, however, carries no information 
on the $T$-dependence of $A^{(0)}_i$'s, which are to be 
determined by the next order equation. The next order 
equation is given by
\begin{eqnarray}
    \dfrac{\rd A^{(1)}_i}{\rd t_j} 
    &=& (1 - \delta_{ij}) 
        \dfrac{[A^{(0)}_i,A^{(1)}_j] + [A^{(1)}_i,A^{(0)}_j]}{T_i - T_j} 
           \nonumber \\
    & & - \delta_{ij} \sum_{k (\not= i)} 
        \dfrac{[A^{(0)}_i,A^{(1)}_k] + [A^{(1)}_i,A^{(0)}_k]}{T_i - T_k}
           \nonumber \\
    & & + (\mbox{terms containing $A^{(0)}$'s and 
          their $T$-derivatives only}). 
\end{eqnarray}
This equation contains $T$-derivatives of $A^{(0)}_i$ as well 
as $t$-derivatives of $A^{(1)}_i$. A standard prescription of 
multiscale analysis is to eliminate the latter by ``averaging 
over the $t$ space.''  This procedure usually takes the form 
of an ``integrability condition'' for the above equation 
to have a solution $A^{(1)}_i$ under suitable a boundary 
condition (e.g., requiring the same quasi-periodicity as 
$A^{(0)}_i$'s possess).  

One thus obtains a differential equation (in $T$) for 
$t$-averaged functionals of $A^{(0)}_i$'s.  This is one 
of various possible expressions (``modulation equations'') 
of modulational dynamics. As Whitham pointed out 
\cite{bib:Whitham}, such a modulation equation frequently 
appears in the form of ``averaged conservation laws.''  
Having this fact in mind, Flaschka, Forest and McLaughlin 
\cite{bib:FFM} considered averaged conservation lows of 
the KdV equation, and derived their compact expression of 
this problem.

\subsection{Our approach to Modulation Equation}

There are many technical difficulties in deriving a 
modulation equation from the Schlesinger equation along 
the line presented above.  A main obstacle is the fact 
that the spectral curve is no longer hyperelliptic for 
$r > 2$.  Most attempts in the literature, including the 
work of Flaschka, Forest and McLaughlin, have been limited 
to hyperelliptic spectral curves. This considerably reduces 
the obstacles (though a complete treatment of the problem 
is still by no means an easy task).  The spectral curve 
of the Schlesinger equation is hyperelliptic only if 
$r = 2$. In the general case, the averaging method is 
inevitably confronted with delicate problems of geometry 
of spectral curves.  (In this respect, Krichever's 
averaging method \cite{bib:Krichever-averaging} 
for general spectral curves is quite remarkable.) 

Since our main concern is the structure of the modulation 
equation rather than the averaging method itself, we now bypass 
these delicate issues by a very heuristic argument.

\section{Modulation Equation} 

\subsection{Multiscale Analysis of Auxiliary Linear Problem}

The first step of our heuristic argument is to apply the 
concept of multiscale analysis to the auxiliary linear 
problem of the Schlesinger equation. Besides the  
multiscale expression of $A_i$'s, we now assume the 
following ansatz to $Y$ (which is now understood to be 
{\it vector-valued}): 
\begin{equation}
    Y = \Bigl( \phi^{(0)}(t,T,\lambda) 
        + \phi^{(1)}(t,T,\lambda) \epsilon + \cdots \Bigr) 
        \exp \epsilon^{-1} S(T,\lambda). 
\end{equation}
Here $\phi^{(k)}(t,T,\lambda)$ are vector-valued function 
and $S(T,\lambda)$ a scalar-valued function. This is a 
kind of JWKB ansatz (inspired by the work of Flaschka and
Newell \cite{bib:FN} and Novikov \cite{bib:Novikov}). 
Note that the ``phase function'' $S$ is independent of $t$. 
We can now write the auxiliary linear problem in the 
following multiscale form: 
\begin{eqnarray}
    \epsilon \frac{\rd Y}{\rd \lambda} 
      = M(\lambda)Y, \\
    \left( \frac{\rd}{\rd t_i} + 
      \epsilon \frac{\rd}{\rd T_i} \right) Y 
      = - \dfrac{A_i}{\lambda - T_i} Y. 
\end{eqnarray}

The leading order equation should reproduce the auxiliary 
linear problem of the isospectral problem. Let us confirm 
that this is indeed the case.  The leading order equation 
is given by 
\begin{eqnarray}
    \frac{\rd S(\lambda)}{\rd \lambda} \phi^{(0)}(\lambda) 
      = M^{(0)}(\lambda) \phi^{(0)}(\lambda), \\
    \frac{\rd \phi^{(0)}(\lambda)}{\rd t_i} 
    + \dfrac{\rd S(\lambda)}{\rd T_i} \phi^{(0)}(\lambda) 
      = - \dfrac{A^{(0)}_i}{\lambda - T_i} \phi^{(0)}, 
\end{eqnarray}
where 
\begin{equation}
    M^{(0)}(\lambda) 
    = \sum_{i=1}^N \dfrac{A^{(0)}_i}{\lambda - T_i}. 
\end{equation}
We now define 
\begin{equation}
    \mu = \dfrac{\rd S(\lambda)}{\rd \lambda}, \quad 
    \psi = \phi^{(0)}(\lambda) 
            \exp \sum_{i=1}^N t_i 
             \frac{\rd S(\lambda)}{\rd T_i}. 
\end{equation}
In terms of these quantities, the leading order equation 
of multiscale expansion can be rewritten 
\begin{equation} 
    \mu \psi = M^{(0)}(\lambda) \psi, \quad 
    \frac{\rd \psi}{\rd t_i} 
             = - \dfrac{A^{(0)}_i}{\lambda - T_i} \psi. 
\end{equation} 
This is exactly the auxiliary linear problem of the 
isospectral problem that we derived in Section 4!

\subsection{Matching with Baker-Akhiezer Function} 

The next step is the most crucial part of our heuristics. 
We now compare the above $\psi$ with the Baker-Akhiezer 
function in the ordinary algebro-geometric approach 
\cite{bib:DMN,bib:Krichever1977} to finite-band solutions 
of soliton equations. 

It is well known that the solution of the auxiliary 
linear problem for a finite-band solution can be 
constructed as a (scalar- or vector-valued) 
Baker-Akhiezer function. In the present setting, 
such a Baker-Akhiezer function can be written 
\begin{equation}
    \psi = \phi \exp \sum_{i=1}^N t_i \Omega_i. 
\end{equation}
Here $\phi$ is a vector-valued function on the spectral 
curve, also depending on $t$; each entry is a combination 
of theta functions.  We do not specify its detailed 
structure, because it is irrelevant in the following 
consideration. Meanwhile, $\Omega_i$ is the primitive 
function of a meromorphic differential $d\Omega_i$ 
on the spectral curve, 
\begin{equation}
    \Omega_i = \int^{(\lambda,\mu)} d\Omega_i. 
\end{equation}
Note that these quantities also depend on  $T$ through 
the $T$-dependence of the spectral curve.  Another 
important remark is that such an expression of the 
Baker-Akhiezer function is available only after 
selecting a ``symplectic homology basis'' of the spectral 
curve, i.e., cycles $A_I,B_I$ ($I = 1,\cdots,g$) with 
intersection numbers $A_I \cdot A_J = B_I \cdot B_J = 0$ 
and $A_I \cdot B_J = \delta_{IJ}$.  (This issue will be 
discussed in detail in Section 6.) 

We now {\it assume} that, for a suitable symplectic 
homology basis, this Baker-Akhiezer function coincides 
with the $\psi$ derived from the leading order equation 
of multiscale expansion.  More precisely, we require 
their ``amplitude part'' and ``phase part,'' respectively, 
to coincide: 
\begin{equation}
    \phi^{(0)} = \phi, \quad 
    \frac{\rd S}{\rd T_i} = \Omega_i. 
\end{equation}
Actually, the first relation may be relaxed as
\begin{equation}
    \phi^{(0)} = \phi h(T,\lambda), 
\end{equation}
where $h(T,\lambda)$ is a scalar function independent 
of $t$; this takes into account the obvious symmetry 
\begin{equation}
    \phi^{(0)} \to \phi^{(0)} h(T,\lambda)
\end{equation}
of the lowest order equation of multiscale expansion. 
Such a factor $h(T,\lambda)$ might be necessary to 
proceed to the next order approximation of multiscale 
analysis.  We shall not go into this issue here. 

\medskip
{\it Remark\/}. 
We take this opportunity to correct an error in the 
previous publication \cite{bib:TN}. Substantially 
the same JWKB ansatz also assumed therein, but with 
an extra factor of the form $(\rd^2 S/\rd \lambda^2)^{1/2}$ 
--- see Eqs. (13) and (21) therein.  This factor 
itself is nonsense.  Fortunately, this does not 
affect the leading order of multiscale analysis.  
As mentioned above, however, something like this 
factor will be necessary in the next order. The 
arbitrary function $h$ stands for such a compensating 
factor.

\subsection{Modulation Equation}

Thus, under several assumptions, we have been able to 
derive a series of relations that link the isomonodromic 
and isospectral problems. In particular, the following 
equations are obtained: 
\begin{equation}
    \frac{\rd S}{\rd \lambda} = \mu, \quad 
    \frac{\rd S}{\rd T_i} = \Omega_i. 
\end{equation}
We propose these equations as the modulation equation 
that governs slow dynamics of the spectral curve 
\begin{equation} 
    \det\Bigl( M^{(0)}(\lambda) - \mu I \Bigr) = 0. 
\end{equation}

Remember that one has to select a symplectic homology 
basis $A_I,B_I$ ($I = 1,\cdots,g$) in order to derive 
these equations. Actually, selecting a proper homology 
basis is a non-trivial problem. We discuss this issue 
in more detail in Section 6, along with a precise 
characterization of the meromorphic differentials $d\Omega_i$.

\subsection{Relation to Generic Whitham Equation} 

As a piece of evidence that our modulation equation is 
a reasonable one, we now show that solutions of this 
equation give a special subfamily of generic Whitham 
deformations. 

The first equation of the modulation equation can be rewritten 
\begin{equation}
    dS = \frac{\rd S}{\rd \lambda} d\lambda = \mu d\lambda. 
\end{equation}
In other words, $S$, like $\Omega_i$'s, is the primitive 
function of a meromorphic differential $dS$ on the 
spectral curve: 
\begin{equation}
    S = \int^{(\lambda,\mu)} \mu d\lambda. 
\end{equation}

The second part of the modulation equation implies the 
equations 
\begin{equation}
    \frac{\rd}{\rd T_i} dS = d\Omega_i. 
\end{equation}
Accordingly, the generic Whitham equation 
\begin{equation}
    \frac{\rd}{\rd T_i} d\Omega_j = \frac{\rd}{\rd T_j} d\Omega_j 
\end{equation}
follows immediately. (Here, as usual, $\lambda$ is understood 
to be {\it constant} under the $T$-derivation. Geometrically, 
this should be treated as a connection.  See the paper of 
Krichever and Phong \cite{bib:Krichever-Phong}.) 

Thus, our modulation equation turns out to yield a special 
subfamily of generic Whitham deformations. We shall see 
in Section 8 that an inverse period map yields generals 
solution of this equation.

\subsection{Possible Approach from Averaging Method}

Our heuristic argument has to be justified in a more 
rigorous form, or at least cross-checked by some other 
method.  A possible check will be to derive the above 
equations by an averaging argument. 

Krichever's averaging method \cite{bib:Krichever-averaging} 
(see also the papers by Fucito et al. \cite{bib:Fucito-etal} 
Carroll and Chang \cite{bib:Carroll-Chang}) seems to 
be particularly promising.  This method employs the 
so called ``dual Baker-Akhiezer function'' $\psi^*$, i.e., 
a solution of the dual auxiliary linear problem 
\begin{equation} 
    \mu \psi^* = \psi^* M^{(0)}(\lambda), \quad 
    \frac{\rd \psi^*}{\rd t_i} 
              = \psi^* \dfrac{A^{(0)}_i}{\lambda - T_i}, 
\end{equation} 
along with $\psi$. One can derive, for instance, the 
following formula evaluating $T$-dependence of the 
eigenvalue $\mu$ of $M^{(0)}$: 
\begin{equation}
    \frac{\rd \mu}{\rd T_i} 
    = \dfrac{\Bigl< \psi^* A^{(0)}(\lambda - T_i)^{-2} \psi \Bigr>}
            {<\psi^* \psi>}. 
\end{equation} 
Here $<\cdots>$ means the average over the fast variables $t$. 
This formula can be readily translated to the language of $dS$ 
(though we have been unable to calculate the average and to 
identify the resulting equation with the above equation for 
$dS$).  Note that the term $A^{(0)}(\lambda - T_i)^{-2}$ in 
the above formula is the same as the ``driving force'' term 
that we encountered in the derivative formula of the characteristic 
polynomial of $M(\lambda)$.

\section{Structure of Meromorphic Differentials}

In this and subsequent sections, we omit the superfix 
``$^{(0)}$'' and write $A^{(0)}_i$, $M^{(0)}(\lambda)$, 
etc. as $A_i$, $M(\lambda)$, $\cdots$.  This is just 
for simplifying notations.

\subsection{General Remarks}

The vector-valued Baker-Akhiezer function $\psi$ is single 
valued on the spectral curve, meromorphic outside the points 
of $\pi^{-1}(\{T_1,\cdots,T_N\})$, and has essential 
singularities of a particular exponential form at these 
exceptional points. The entries of $\phi$ have a common 
pole divisor $D$, part of which may overlap with the 
essential singularities. (In the following, we consider 
a generic case where this overlapping does not occur.)  
The essential singularities at points in $\pi^{-1}(T_i)$ 
are generated by poles of the meromorphic differential 
$d\Omega_i$. 

The meromorphic differentials $d\Omega_i$ have to be 
selected for the essential singularities of $\psi$ to 
match the auxiliary linear problem. This yields conditions 
on the poles of the meromorphic differentials. Our task 
in the following is to specify those conditions. 

In order to consider this problem, it is convenient to build 
a matrix solution $\Psi$ of the auxiliary linear problem 
from the vector-valued Baker-Akhiezer function $\psi$. Let 
$P_\alpha(\lambda)$ ($\alpha = 1,\cdots,r$), denote the points 
of $\pi^{-1}(\lambda)$.  (Of course, also here, such numbering 
is meaningful only locally.)  $\Psi$ is given by 
\begin{equation}
    \Psi = \Bigl( \psi(P_1(\lambda)),\cdots,\psi(P_r(\lambda)) \Bigr). 
\end{equation}
This matrix has the following factorized form: 
\begin{equation}
    \Psi = \Phi \exp \diag\Bigl( 
            \sum t_i \Omega_i( P_1(\lambda) ), \cdots, 
            \sum t_i \Omega_i( P_r(\lambda) ) \Bigr). 
\end{equation}
$\Phi = \Phi(\lambda)$ is an $r \times r$ matrix originating 
in the ``amplitude part'' $\phi$, and invertible at a generic 
point.

\subsection{Conditions on Meromorphic Differentials}

\subsubsection{Poles of $d\Omega_i$}

All necessary information on poles of $d\Omega_i$ can be 
derived from the following relation, which is obtained by 
inserting the above expression of $\Psi$ into the auxiliary 
linear problem: 
\begin{equation}
    \frac{\rd \Phi}{\rd t_i} \Phi^{-1} 
    + \Phi \diag\Bigl( 
        \Omega_i( P_1(\lambda) ), \cdots, 
        \Omega_i( P_r(\lambda) ) \Bigr) \Phi^{-1}  
    = - \dfrac{A_i}{\lambda - T_i}. 
\end{equation}

We now compare the singular part of both hand side at 
$\lambda = T_i$. Since the first term on the left hand side 
is non-singular, it turns out that $\Omega_i$ has a pole of 
first order at each point of $\pi^{-1}(T_i)$, and the residue 
is equal to $(-1)$ times an eigenvalues of $A_i$. Since the 
eigenvalues of $A_i$ are $\theta_{i\alpha}$, this implies that
(after suitably renumbering the eigenvalues)  $A_i$ can be written 
\begin{equation}
    A_i = \Phi(T_i) \diag\Bigl( \theta_{i1},\cdots,\theta_{ir} \Bigr) 
          \Phi(T_i)^{-1}, 
\end{equation}
and that 
\begin{equation}
    \Omega_i = - \dfrac{\theta_{i\alpha}}{\lambda - T_i} 
        + \mbox{non-singular} 
\end{equation}
in a neighborhood of $P_\alpha(T_i)$. 

In particular, the meromorphic differential $d\Omega_i$ 
behaves as
\begin{equation}
    d\Omega_i 
    = \dfrac{\theta_{i\alpha}}{(\lambda - T_i)^2} d\lambda 
      + \mbox{non-singular}
\end{equation}
in the same neighborhood of $P_\alpha(T_i)$. Therefore, 
if $\theta_{i\alpha} \not= 0$, $P_\alpha(T_i)$ is a pole 
of $d\Omega_i$ of second order.  These are all poles that 
$d\Omega_i$ is required to have.

\subsubsection{Period Integrals of $d\Omega_i$}

The above conditions on poles determine the meromorphic 
differential $d\Omega_i$ up to a difference of holomorphic 
differential. We now select a symplectic homology basis 
$A_I,B_I$ ($I = 1,\dots,g$), and put the standard 
normalization condition
\begin{equation}
    \oint_{A_I} d\Omega_i = 0, \quad I = 1,\dots,g. 
\end{equation}
$d\Omega_i$ is thus uniquely determined.

\subsection{Back to Modulation Equation} 

This is the end of the precise description of our modulation 
equation.  Let us finally reconfirm the roles of the three 
sets of parameters $\theta_{i\alpha}$, $T_i$ and $h_{m\ell}$ 
in the defining equation of the spectral curve: 
\begin{itemize}
\item The coadjoint orbit invariants $\theta_{i\alpha}$ are 
      constant parameters. 
\item The positions of poles $T_i$ are ``time variables.''
\item The isospectral invariants $h_{m\ell}$ are ``dynamical 
      variables.''
\end{itemize}

\subsection{Remarks on Symplectic Homology Basis} 

The choice of symplectic homology basis is a non-trivial 
problem.  The work of Flaschka, Forest and McLaughlin 
\cite{bib:FFM} provides a typical example for considering 
this problem.  In their work, the meromorphic differentials 
$d\Omega_i$ are normalized along the so called ``$\mu$-cycles.'' 
These cycles are homologous to trajectories of the 
``auxiliary spectrum,'' which are nothing but the degree 
$g$ divisor $\sum_{i=1}^g (\lambda_i,\zeta_i)$ of zeros of 
a Baker-Akhiezer function on the KdV hyperelliptic spectral 
curve 
\begin{equation}
    \zeta^2 = \prod_{i=1}^{2g+1} (\lambda - e_i). 
\end{equation}
The Whitham averaging over the quasi-periodic motion on 
the Jacobian variety eventually boils down to period 
integrals along these trajectories.  As this typical 
example shows clearly, the normalization condition 
of $d\Omega_i$ is by no means arbitrary, but determined 
by the geometric structure of motion of a divisor on 
the spectral curve. (In fact, as Ercolani, Forest and 
McLaughlin noted \cite{bib:EFM}, this issue is already 
considerably delicate in the case of the sine-Gordon 
equation.) 

Meanwhile, our formulation of the modulation equation 
itself works for any choice of the symplectic homology 
basis. Furthermore, the results of the subsequent 
sections also hold irrespective of the choice of the 
symplectic homology basis. Presumably, it is only a 
subset of solutions of our modulation equation that 
actually correspond to the true modulational description 
of isomonodromic problems. This issue, too, forms 
part of the hard analytical problems that we do not 
pursue in this paper.

\section{Solutions of Modulation Equation}

\subsection{Period Integrals \`a la Seiberg-Witten} 

In the following, let $h = \{ h_I \mid \ I=1,\cdots,g \}$ 
denote  the $g$-tuple of isospectral invariants $h_{m\ell}$ 
($\ell = 2,\cdots,r$, $m = 0,\cdots,\delta_\ell - 1$) 
(see Section 3). With the other parameters $\theta_{i\alpha}$ 
and $T_i$ being fixed, the spectral curve forms a $g$-dimensional 
deformation family parametrized by these isospectral invariants.  
This is the same situation as the Seiberg-Witten solution and 
its various generalizations \cite{bib:SW-integrable}.  One can 
indeed derive the following analogous results. 
\begin{itemize}
\item 
  A set of (local) coordinates on the $g$-dimensional moduli space 
  of spectral curves are given by the period integrals
  \begin{equation}
    a_I = \oint_{A_I} dS, \quad I = 1,\cdots,g. 
  \end{equation}
  Here (and in the following),  $ A_I, B_I$ ($I = 1,\cdots,g$) 
  are the same symplectic homology basis as used in the definition 
  of the modulation equation. 
\item
  Another set of (local) coordinates are given by the dual period 
  integrals
  \begin{equation}
    b_I = \oint_{B_I} dS, \quad I = 1,\cdots,g. 
  \end{equation}
  In the Seiberg-Witten theory, they are denoted by $a^D_I$. 
\item 
  There is a (locally defined) function $\calF = \calF(a)$ 
  (``prepotential'') of $a = \{ a_I \}$  such that 
  \begin{equation}
    \frac{\rd \calF}{\rd a_I} = b_I. 
  \end{equation}
\item 
  The second derivatives of $\calF$ coincide with matrix elements 
  of the period matrix, 
  \begin{equation}
    \frac{\rd^2 \calF}{\rd a_I \rd a_J} = \calT_{IJ} 
     = \oint_{B_I} d\omega_J,     
  \end{equation}
  where $d\omega_I$ ($I = 1,\cdots,g$) are a basis of holomorphic 
  differentials uniquely determined by the normalization condition
  \begin{equation}
    \oint_{A_I} d\omega_J = \delta_{IJ}.
  \end{equation}
\end{itemize}
In a sense, the rest of this paper is devoted to verifying 
these results in an $T$-dependent form.  

The goal of this section is to show that the inverse period 
map $a \mapsto h$ solves our deformation equation. We first 
establish the invertibility of the period map $h \mapsto a$. 
The inverse map $a \mapsto h$ then turns out to satisfy a 
deformation equation of Seiberg-Witten type (with respect 
to $a_i$'s), as well as our modulation equation (with respect 
to $T_i$'s).  Most results and proofs are a rather straight 
forward generalization of those of the Seiberg-Witten 
solution \cite{bib:SW-integrable}. 

The notion of prepotential will be discussed in detail in 
Section 8.

\subsection{Invertibility of Period Map} 

This subsection is organized as follows. Firstly, we construct 
a basis $d\omegatilde_I$ ($I = 1,\cdots,g$) of holomorphic 
differentials as derivatives of $dS$ with respect to $h_I$'s. 
Secondly, we examine the linear relations between this basis 
and the normalized basis $d\omega_I$ ($I = 1,\cdots,g$). As a 
corollary, (local) invertibility of the period map $h \to a$ 
follows.

\subsubsection{Basis of Holomorphic Differentials}

We here consider $\mu$ to be a (multivalued) function 
$\mu = \mu(T,h,\lambda)$ of $T$, $h$ and $\lambda$, and 
differentiate $dS(T,h,\lambda) = \mu(T,h,\lambda) d\lambda$
by $h_I$. This gives the differential 
\begin{equation}
    d\omegatilde_I = \frac{\rd}{\rd h_I} dS 
    = \frac{\rd \mu(T,h,\lambda)}{\rd h_I} d\lambda. 
\end{equation}
We now show that this gives a basis of holomorphic differentials 
on the spectral curve. The following reasoning is borrowed 
from the work of Adams, Harnad and Hurtubise \cite{bib:AHH}. 

This definition of $d\omegatilde_I$ can be cast into a more 
tractable form as follows. First rewrite it in terms of 
$\mutilde = f(\lambda) \mu$ (see Section 3): 
\[ 
    d\omegatilde_I 
    = \frac{1}{f(\lambda)} 
        \frac{\rd \mutilde(T,h,\lambda)}{\rd h_I} d\lambda. 
\] 
Differentiating the equation 
\[ 
    \Ftilde(\lambda,\mutilde) 
    = \det \Bigl( f(\lambda)M(\lambda) - \mu I \Bigr) = 0 
\] 
of the spectral curve in the $(\lambda,\mutilde)$ coordinates 
gives 
\[
    \frac{\rd \mutilde}{\rd h_I} 
      = - \dfrac{\rd \Ftilde(\lambda,\mutilde)/\rd h_I}
                {\rd \Ftilde(\lambda,\mutilde)/\rd \mutilde}. 
\]
Now recall that $h_I$ is just an abbreviation of one of 
$h_{m\ell}$'s in the following formula of Section 3: 
\[
    \Ftilde(\lambda,\mutilde) = 
      \Ftilde^0(\lambda,\mutilde) 
      + f(\lambda) \sum_{\ell=2}^r \sum_{m=0}^{\delta_\ell} 
          h_{m\ell} \lambda^m \mutilde^{r-\ell}. 
\]
Therefore, for $h_I = h_{m\ell}$, 
\[
    \frac{\rd \Ftilde(\lambda,\mutilde)}{\rd h_{m\ell}} 
    = f(\lambda) \lambda^\ell \mutilde^{r-m}. 
\]
Thus, eventually, we obtain the following expression 
of $d\omegatilde_I = d\omegatilde_{m\ell}$: 
\begin{equation}
    d\omegatilde_{m\ell} 
    = - \dfrac{\lambda^\ell \mutilde^{r-m}}
        {\rd \Ftilde(\lambda,\mutilde)/\rd \mu} d\lambda. 
\end{equation}

The last expression may be viewed as the Poincar\'e residue 
of a meromorphic 2-form on the $(\lambda,\mutilde)$ plane 
with pole divisor along the spectral curve.  Since 
$2 \le \ell \le r$ and $0 \le m \le \delta_\ell - 1$, 
a standard argument leads to the following result: 

\begin{proposition}
The differentials $d\omegatilde_I$ ($I = 1,\cdots,g$), form 
a basis of holomorphic differentials on the spectral curve. 
\end{proposition}

\subsubsection{Linear Relations between Two Bases} 

Since $d\omegatilde_I$ and $d\omega_I$ both give a basis of 
holomorphic differentials, they should be linked by an 
invertible linear transformation: 
\begin{equation}
    d\omegatilde_I = \sum_{J=1}^g \calA_{IJ} d\omega_J. 
\end{equation}
The matrix $\calA = (\calA_{IJ})$ of the coefficients is 
invertible.  

In fact, the matrix $\calA$ is given by the period integrals 
\begin{equation}
    \calA_{IJ} = \oint_{A_J} d\omegatilde_I, 
\end{equation}
as one can readily see by integrating the above linear relation 
of $d\omegatilde_I$ and $d\omega_J$ along $A_J$. Similarly, 
integrating along $B_J$ yields the matrix relation 
\begin{equation}
    \calB = \calA \calT, 
\end{equation}
where the matrix elements of $\calB = (\calB_{IJ})$ are given by  
\begin{equation}
    \calB_{IJ} = \oint_{B_J} d\omegatilde_I.  
\end{equation}
In particular, the period matrix $\calT$ can be written 
\begin{equation} 
    \calT = \calA^{-1} \calB. 
\end{equation}

\subsubsection{Invertibility of Period Map}

We now show the following result, which shows that the matrix 
is in fact the Jacobian matrix of the period map $h \mapsto a$. 

\begin{proposition}
\begin{equation}
    \calA_{IJ} = \dfrac{\rd a_J}{\rd h_I}. 
\end{equation}
\end{proposition}

{\it Proof\/}. $a_J$ is defined by the period integral 
\[
    a_J = \oint_{A_J} dS. 
\]
Its $h_I$-derivative can be calculates as follows: 
\[
    \dfrac{\rd a_J}{\rd h_I} 
    = \oint_{A_J} \dfrac{\rd}{\rd h_I} dS 
    = \oint_{A_J} d\omegatilde_I 
    = \calA_{IJ}. 
\]
Q.E.D. 

Since $\calA$ is an invertible matrix, we have: 

\begin{corollary} The period map $h \mapsto a$ is (locally) 
invertible. 
\end{corollary}

\subsection{Solving Deformation Equations by Inverse Period Map} 

We now take into account the variables $T = \{T_i\}$, 
and consider the $I$-th coordinate of the inverse map 
$a \mapsto h$ as a function of $T$ and $a$: 
\begin{equation}
    h_I = h_I(T,a).
\end{equation}
(In order to obtain these functions, one has to solve 
the defining equation of the $a_I$'s for the parameters 
$h_I$. This is by no means an easy task.)  Accordingly, 
$dS = \mu d\lambda$ becomes a meromorphic differential 
depending on the parameters $(T,a)$: 
\begin{equation}
    dS = dS(T,a). 
\end{equation}
We can now prove that $dS$ satisfies the modulation equation 
with respect to $T$: 

\begin{proposition}
$dS$ satisfies the modulation equation 
\[
    \frac{\rd}{\rd T_i} dS = d\Omega_i.
\]
\end{proposition}

{\it Proof\/}. Let us tentatively define 
\[
    d\Omegatilde_i = \frac{\rd}{\rd T_i} dS 
    = \frac{\rd \mu}{\rd T_i} d\lambda.
\]
We show that this differential satisfies all conditions 
that characterize $d\Omega_i$. By the uniqueness, then, 
$d\Omegatilde_i$ coincides with $d\Omega_i$, and the 
proof is completed. 

(i) {\it Locations of poles of $d\Omegatilde_i$\/}. 
Differentiating the equation of the spectral curve
\[
    F(\lambda,\mu) = \det \Bigl( M(\lambda) - \mu I \Bigr) = 0 
\]
gives the relation
\[
    \frac{\rd \mu}{\rd T_i} 
      = - \dfrac{\rd F(\lambda,\mu)/\rd T_i}
                {\rd F(\lambda,\mu)/\rd \mu}. 
\]
Therefore, apart from the poles of $\mu$, the derivative 
$\rd \mu/\rd T_i$ can have poles at the zeros of 
$\rd F/\rd \mu$ (i.e., at ramified points of $\pi$). 
Let $e$ be the ramification index at a ramified point.  
$\rd \mu/\rd T_i$ has a pole of order at most $e-1$ there. 
This pole, however, is canceled by zeros of $d\lambda$ 
because $d\lambda$ has a zero of order $e - 1$ at the 
same point.  Meanwhile, in a neighborhood of each point of 
$\pi^{-1}(\infty)$, 
\[
    \frac{\rd \mu}{\rd T_i}
     = \left( \theta_{\infty\alpha} \lambda^{-2} 
        + O(\lambda^{-2}) \right) d\lambda, 
\]
so that the derivative $\rd \mu/\rd T_i$ cancels the 
second order poles of $d\lambda$ at these points. 
Thus, poles of $d\Omegatilde_i$ are limited to points 
in $\pi^{-1}(\{T_1,\cdots,T_N\})$. 

(ii) {\it Singular behavior of $d\Omegatilde_i$ at poles\/}. 
We examine the singular behavior of $d\Omegatilde_i$ at each point 
of $\pi^{-1}(T_j) = \{P_{j1},\cdots,P_{jr}\}$. Recall that, 
in a neighborhood of $P_{i\alpha}$, $\mu$ behaves as
\[
    \mu = - \dfrac{\theta_{i\alpha}}{\lambda - T_j} 
          + \mbox{non-singular}. 
\]
Therefore 
\[
    \frac{\rd \mu}{\rd T_i} = \left\{ \begin{array}{ll}
          \mbox{non-singular} & (j \not= i) \\
          - \dfrac{\theta_{i\alpha}}{(\lambda - T_i)^2} 
          + \mbox{non-singular} & (j = i)
        \end{array} \right.
\]
Consequently, $d\Omegatilde_i$ has poles only at the points over 
$\lambda = T_i$, and exhibits there the same singular 
behavior as $d\Omega_i$ does. 

(iii) {\it Period Integrals of $d\Omegatilde_i$\/}. Since 
the deformations leave $a_I$'s invariant, we have 
\[
    \oint_{A_I} \frac{\rd}{\rd T_i} dS 
    = \frac{\rd}{\rd T_i} \oint_{A_I} dS 
    = \frac{\rd a_I}{\rd T_i} = 0. 
\]
Thus $d\Omegatilde_i$ satisfies the normalization condition of periods, too. 
Q.E.D.

Similarly, $dS$ turns out to satisfy a deformation equation of 
Seiberg-Witten type with respect to $a_I$'s: 

\begin{proposition} 
$dS$ satisfies the deformation equation 
\[
    \frac{\rd}{\rd a_I} dS = d\omega_I, \quad I = 1,\cdots,g, 
\]
of Seiberg-Witten type.
\end{proposition} 

{\it Proof\/}. This is just a consequence of the chain rule 
of differentiation: 
\[
    \frac{\rd}{\rd a_I} dS 
    = \sum_J \frac{\rd h_J}{\rd a_I} \frac{\rd}{\rd h_J} dS 
    = \sum_J (\calA^{-1})_{IJ} d\omegatilde_J 
    = d\omega_I. 
\]
Q.E.D. 

Thus, we have been able to show that the inverse period map 
$a \mapsto h$ solves the coupled deformation equations 
\begin{equation}
    \frac{\rd}{\rd T_i} dS = d\Omega_i, \quad 
    \frac{\rd}{\rd a_I} dS = d\omega_I. 
\end{equation}
(Reconfirm, once again, that the differentiation is 
understood to leave $\lambda$ constant, i.e., 
$\rd \lambda/\rd T_i =  \rd \lambda/\rd a_I = 0$.)

\subsection{Remarks on Structure of $dS$}

It seems remarkable that although our meromorphic 
differential $dS$ has simple poles with non-zero residues, 
their derivatives in deformation variables are holomorphic 
differentials or meromorphic differentials of second kind 
(i.e., meromorphic differentials with higher order poles 
and no residue). This is reminiscent of the structure of 
the Seiberg-Witten meromorphic differential in 
supersymmetric gauge theories coupled with matter fields 
\cite{bib:SW-integrable}. 

Because of this fact, $dS$ cannot be written as a linear
combination of $d\omega_I$ and $d\Omega_i$; we need extra 
meromorphic differentials of the third kind, say $d\Pi$, 
with the same singularity structure as $dS$ but with 
vanishing $A_I$-cycles. $dS$ can be written 
\begin{equation}
    dS = \sum_{I=1}^g a_I d\omega_I + d\Pi. 
\end{equation}
Of course $d\Pi$ can be further decomposed into a sum of 
more elementary meromorphic differentials, but can never 
be a linear combination of $d\Omega_i$'s.

\section{Prepotential}

\subsection{Definition of Prepotential}

The notion of prepotential has been formulated in a quite 
general framework by Krichever \cite{bib:Krichever-Whitham}
and Dubrovin \cite{bib:Dubrovin-Whitham}. 
Following their formalism we can now define a prepotential 
$\calF = \calF(T,a)$ by the following differential equation.

\begin{proposition}
The following differential equations is integrable in the 
sense of Frobenius: 
\begin{equation}
    \frac{\rd \calF}{\rd a_I} = \oint_{B_I} dS, \quad 
    \frac{\rd \calF}{\rd T_i} 
      = \sum_{\alpha=1}^r \Res_{P_{i\alpha}} 
        \dfrac{\theta_{i\alpha}}{\lambda - T_i} dS. 
\end{equation}
Here $P_{i\alpha}$, $\alpha = 1,\cdots,r$, denote the points 
of $\pi^{-1}(T_i)$ such that 
\begin{equation}
    \mu = \dfrac{\theta_{i\alpha}}{\lambda - T_i} 
          + \mbox{non-singular} 
\end{equation}
in a neighborhood of $P_{i\alpha}$.  
\end{proposition}

\subsection{Equivalent Definition} 

Before proving the above result, we note here that the 
defining equations of $\calF$ can be rewritten in the 
following more compact form: 
\begin{equation}
    \frac{\rd \calF}{\rd a_I} = b_I, \quad 
    \frac{\rd \calF}{\rd T_i} = H_i. 
\end{equation}

The equivalence of the first equation is obvious from the 
definition of $b_I$. Let us verify the equivalence of the 
second equation in some detail. We first rewrite the 
definition of $H_i$ into a contour integral  
\[
    H_i = \frac{1}{2\pi\sqrt{-1}} 
          \oint_{|\lambda - T_i| = \delta} 
          \frac{1}{2} \Tr M(\lambda)^2 d\lambda 
\]
along a sufficiently small circle (of radius $\delta$) 
around $\lambda = T_i$. Let $\mu_\alpha$, 
$\alpha = 1,\cdots,r$, denote the eigenvalues of 
$M(\lambda)$; they correspond to the $r$ sheets of 
the spectral curve in a neighborhood of $\pi^{-1}(T_i)$. 
Expressing the trace of $M(\lambda)^2$ in terms of 
these eigenvalues, we can rewrite the above integral 
formula as:
\begin{eqnarray*}
    H_i 
    &=& \frac{1}{2\pi\sqrt{-1}}\oint_{|\lambda - T_i| = \delta} 
        \sum_{\alpha=1}^r \frac{\mu_\alpha^2}{2} d\lambda     \\
    &=& \sum_{\alpha=1}^r 
        \frac{1}{2\pi\sqrt{-1}}\oint_{C_{i\alpha}} 
        \frac{\mu^2}{2} d\lambda. 
\end{eqnarray*}
Now recall that $\mu$ has a Laurent expansion of the form 
\[
    \mu = \dfrac{\theta_{i\alpha}}{\lambda - T_i} 
          + c_{i\alpha,0} + c_{i\alpha,1}(\lambda - T_i) 
          + \cdots 
\]
in a neighborhood of $P_{i\alpha}$. From this fact, 
we can easily derive the relation 
\[
      \oint_{C_{i\alpha}} \frac{\mu^2}{2} d\lambda 
    = \oint_{C_{i\alpha}} 
      \dfrac{\theta_{i\alpha}}{\lambda - T_i} dS. 
\]
This relation and the above expression of $H_i$ give
\[
    H_i = \sum_{\alpha=1}^r \Res_{P_{i\alpha}} 
            \dfrac{\theta_{i\alpha}}{\lambda - T_i} dS. 
\]
This leads to the second expression of the defining 
equation of the prepotential.

\subsection{Proof of Proposition}

The idea of proof is the same as the standard one based on 
Riemann's bilinear relation.  Integrability conditions to be 
checked are the following: 
\begin{equation}
    \dfrac{\rd H_j}{\rd T_i} = \dfrac{\rd H_i}{\rd T_j}, \quad 
    \dfrac{\rd H_j}{\rd a_I} = \dfrac{\rd b_I}{\rd T_j}, \quad
    \dfrac{\rd b_J}{\rd a_I} = \dfrac{\rd b_I}{\rd a_J}. 
\end{equation}
Here $b_I$, $b_J$, $H_i$ and $H_j$ are used for simplifying  
notations; in the verification below, we have to return to the 
integral expressions of these quantities. 

These three relations can be verified in much the same way. 
We show the derivation of the second relation in detail. 
First, differentiating the integral formula 
\[
    H_j = \sum_\alpha \frac{1}{2\pi\sqrt{-1}} \oint_{C_{j\alpha}} 
            \dfrac{\theta_{j\alpha}}{\lambda - T_j} dS 
\]
by $a_I$ gives 
\[
    \dfrac{\rd H_j}{\rd a_I} 
    = \sum_\alpha \frac{1}{2\pi\sqrt{-1}} \oint_{C_{j\alpha}}
        \dfrac{\theta_{j\alpha}}{\lambda - T_j} d\omega_I. 
\]
Similarly, from the integral formula of $b_I$, 
\[
    \dfrac{\rd b_I}{\rd T_j} 
    = \oint_{B_I} d\Omega_j. 
\]
By Riemann's bilinear relation, $d\Omega_j$ and $d\omega_I$ 
satisfies the relation
\[
    (*) \quad 
    \frac{1}{2\pi\sqrt{-1}} \oint_{\rd\Delta} \Omega_j d\omega_I 
    = \sum_J \left( 
        \oint_{A_K} d\Omega_j \oint_{B_K} d\omega_I - 
        \oint_{B_K} d\Omega_j \oint_{A_K} d\omega_I \right). 
\]
Here $\Delta$ is a simply connected surface with boundary 
($4g$-gon) obtained in a standard way by cutting the Riemann 
surface of the spectral curve along $2g$ paths. The boundary 
$\rd\Delta$ is oriented in the direction encircling interior 
points anti-clockwise.  We now evaluate both hand sides. 
As for the left hand side of $(*)$, $\Omega_j$ has poles at 
the points $P_{j1},\cdots,P_{jr}$ in $\Delta$ whereas 
$d\omega_I$ has of course no pole. Therefore the contour 
integral along $\rd\Delta$ splits into a sum of contour 
integrals along $C_{j1},\cdots,C_{jr}$. Furthermore, the 
singular behavior of $\Omega_j$ in a neighborhood of 
$P_{j\alpha}$ is such that 
\[
    \Omega_j = - \theta_{j\alpha}/(\lambda - T_j) 
               + \mbox{non-singular}. 
\]
Therefore 
\[
    \mbox{LHS of $(*)$} 
    = - \sum_\alpha \frac{1}{2\pi\sqrt{-1}} \oint 
      \dfrac{\theta_{j\alpha}}{\lambda - T_j} d\omega_I. 
\]
Now consider the the right hand side of $(*)$. The integrals 
of $d\Omega_j$ along $A_K$'s 
all vanish by the normalization conditions of $d\Omega_j$. 
Its integrals along $B_K$'s do not vanish in general, but 
now the integral of $d\omega_I$ along $A_K$'s vanish 
except for $K = I$; in the case of $K = I$, the latter 
integral is equal to $1$ by the normalization conditions 
of $d\omega_I$. Thus 
\[
    \mbox{RHS of $(*)$} = - \oint_{B_I} d\Omega_j. 
\]
These relations imply the second integrability condition. 

In the verification of the first integrability condition, 
it is convenient to express $d\Omega_i$ as a sum of 
meromorphic differentials $d\Omega_{i\alpha}$ with just 
one pole at $P_{i\alpha}$ and normalized by the same 
condition of vanishing $A$-periods.  The integrability 
condition can then be reduced to Riemann's bilinear 
relation for $d\Omega_{i\alpha}$ and $d\Omega_{j\beta}$. 

The third integrability condition is the easiest to 
verify.  This case, however, is the most interesting 
in the context of the Seiberg-Witten theory. 
Differentiating $b_J$ now by $a_I$ gives 
\[
    \dfrac{\rd b_J}{\rd a_I} = \oint_{B_J} d\omega_I 
    = \calT_{IJ}. 
\]
This is equal to $\rd b_I/\rd a_J$, because the period 
matrix $\calT$ is symmetric. (One can also directly 
deduce this conclusion by applying Riemann's bilinear 
relation to $d\omega_I$ and $d\omega_J$.  This is 
indeed a usual proof of the relation 
$\calT_{IJ} = \calT_{JI}$ !)  

This completes the proof of the proposition.  

As a corollary of the proposition and the final part of 
the proof, we obtain the following result: 
\begin{corollary}
The second $a$-derivatives of $\calF = \calF(T,a)$ 
coincide with the matrix elements of the period matrix:
\begin{equation}
    \dfrac{\rd^2 \calF}{\rd a_I \rd a_J} = \calT_{IJ}. 
\end{equation}
\end{corollary}
Thus our treatment of the prepotential incorporates 
all essential aspects of the prepotential of 
Seiberg-Witten type.

\subsection{Final Remarks} 

(i)  The definition of prepotential shows a link with the 
notion of $\tau$ function. Recall that $H_i$ is equal to 
a logarithmic derivative of the $\tau$ function of the 
Schlesinger equation.  In the $\epsilon$-dependent 
formulation, 
\begin{equation}
    H_i = \epsilon \frac{\rd \log \tau}{\rd T_i}.  
\end{equation}
Note, however, that $\log\tau$ and $\calF$ are by no means 
in a simple proportional relation, because some averaging 
operation intervenes. 

(ii) Unlike the prepotentials in topological conformal 
field theories and the Seiberg-Witten theory, 
our prepotential $\calF$ seems to possess no manifest 
homogeneity of degree two. This issue seems to be 
linked with the unusual structure of $dS$ that 
we pointed out in the end of the last section. 
Presumably our prepotential (and the modulation 
equation) will have hidden homogeneity, which emerges 
after introducing more deformation variables.

\section{Conclusion} 

The body of this paper consists of two part. The 
first part is concerned with the derivation of the 
modulation equation. This equation is expected to 
describe slow dynamics of the spectral curve in  
isospectral approximation, in the sense of Garnier, 
of the ($\epsilon$-dependent) Schlesinger equation. 
Although our method for deriving this modulation 
equation is heuristic, we believe that this gives 
a correct answer (under a suitable choice of the 
symplectic homology basis).  The second part of 
this paper is devoted to a complete description of 
solutions of the modulation equation, as well as the 
notion of prepotential.  We have been able to obtain 
the following fundamental results: 
\begin{itemize}
\item General solutions are obtained by an inverse period 
      map. The period map is given by period integrals of 
      Seiberg-Witten type. 
\item The modulation equation can coexist with a deformation 
      equation of Seiberg-Witten type. They altogether form a 
      commuting set of flows on the $g$-dimensional moduli 
      space of spectral curves. 
\item A prepotential can be defined on this extended commuting 
      flows. This construction is parallel to already known 
      examples of prepotentials. 
\end{itemize}
These results, too, strongly support the validity of the 
modulation equation. 

A number of problems are left unanswered. The most crucial 
is of course the issue of validity of our derivation of 
modulation equation. We have presented a few fragmental 
ideas in this direction.  Krichever's averaging method 
seems to be the most powerful and universal approach to 
this problem; we,  however, have no idea how to calculate 
the averaged quantities to obtain period integrals. 
Presumably, Vereschagin's method will be a  hint to this 
issue.  Also, we would like to stress that generalizing 
the averaging calculation of Flaschka, Forest and 
McLaughlin to non-hyperelliptic spectral curves is still 
a very important problem; the spectral Darboux coordinates 
of the Montreal group should be used in place of the 
classical auxiliary spectrum.

The methods presented in this paper will be generalized 
to other types of isomonodromic problems. An immediate, but 
also interesting generalization is the case with irregular 
singular points.  For instance, if an irregular singular 
point of Poincar\'e rank one is added to $\lambda = \infty$, 
the resulting isomonodromic problem becomes the so called 
JMMS equation \cite{bib:JMMS}. This equations exhibits a 
remarkable ``duality'' \cite{bib:Harnad-dual}, which will 
be inherited to the modulation equation. This issue will 
be discussed in a separate paper \cite{bib:takasaki-dual}. 
An even more interesting direction is a generalization 
to isomonodromic problems on an elliptic (and higher genus) 
Riemann surface.  Examples of such isomonodromic problems 
have been constructed by Okamoto's group in a geometric 
framework \cite{bib:nonzero-genus}.  Recently, Levin and 
Olshanetsky presented another framework based on the 
method of Hamiltonian reduction, and pointed out a link 
with the Whitham equation \cite{bib:levin-olshanetsky}.

\end{document}